\documentclass[conference]{IEEEtran}
\IEEEoverridecommandlockouts
\usepackage{cite}
\usepackage{amsmath,amssymb,amsfonts}
\usepackage{graphicx}
\usepackage{booktabs}
\usepackage{array}
\usepackage{balance}
\usepackage{url}
\usepackage[dvipsnames,svgnames,x11names]{xcolor}
\usepackage{tikz}
\usetikzlibrary{arrows.meta,positioning}
\usepackage[hidelinks]{hyperref}
\usepackage{wasysym}
\usepackage{color}
\usepackage{soul}
\usepackage{multicol}
\usepackage{subcaption}
\usepackage{float}
\usepackage{paralist}

\usepackage[colorinlistoftodos,textsize=footnotesize]{todonotes}
\usepackage{soul}

\graphicspath{{figures/}}

\begin{document}

\title{Reputation as Community Memory \\for the Agentic Web}

\author{
\IEEEauthorblockN{Ryan Chard,
Gus Ellerm,
Alexander Brace,
Alok Kamatar,
Suman Raj,
Ian Foster,
and Kyle Chard}
\IEEEauthorblockA{University of Chicago \& Argonne National Laboratory, Chicago, IL, USA}
}

\maketitle

\begin{abstract}
Agents can now externalize experience into memory, consolidating historical traces into semantic knowledge and procedural shortcuts that persist between sessions. Such memory is typically private to a single agent. We argue that 
agentic memory benefits from being \emph{collective}, because trustworthy knowledge of the shared environment---the data sources, services, and tools agents depend on---cannot be established by any single agent, only corroborated across many independent observers.
We present Cairn, a community reputation platform that captures collective knowledge, allowing
agents to query the community's opinion of a resource before use and to submit evidence-backed ratings afterward. 
Cairn aggregates observations via a time-decayed Beta model with confidence shrinkage and supports semantic discovery over reviewer rationales.
We evaluate Cairn's reputation engine under adversarial simulation (e.g., lying, collusion, camouflage), benchmark its retrieval performance, and report a case study of rating heterogeneous agents in production.

\end{abstract}


\section{Introduction}\label{sec:intro}

\begin{figure*}[t]
    \centering
    \includegraphics[width=\linewidth]{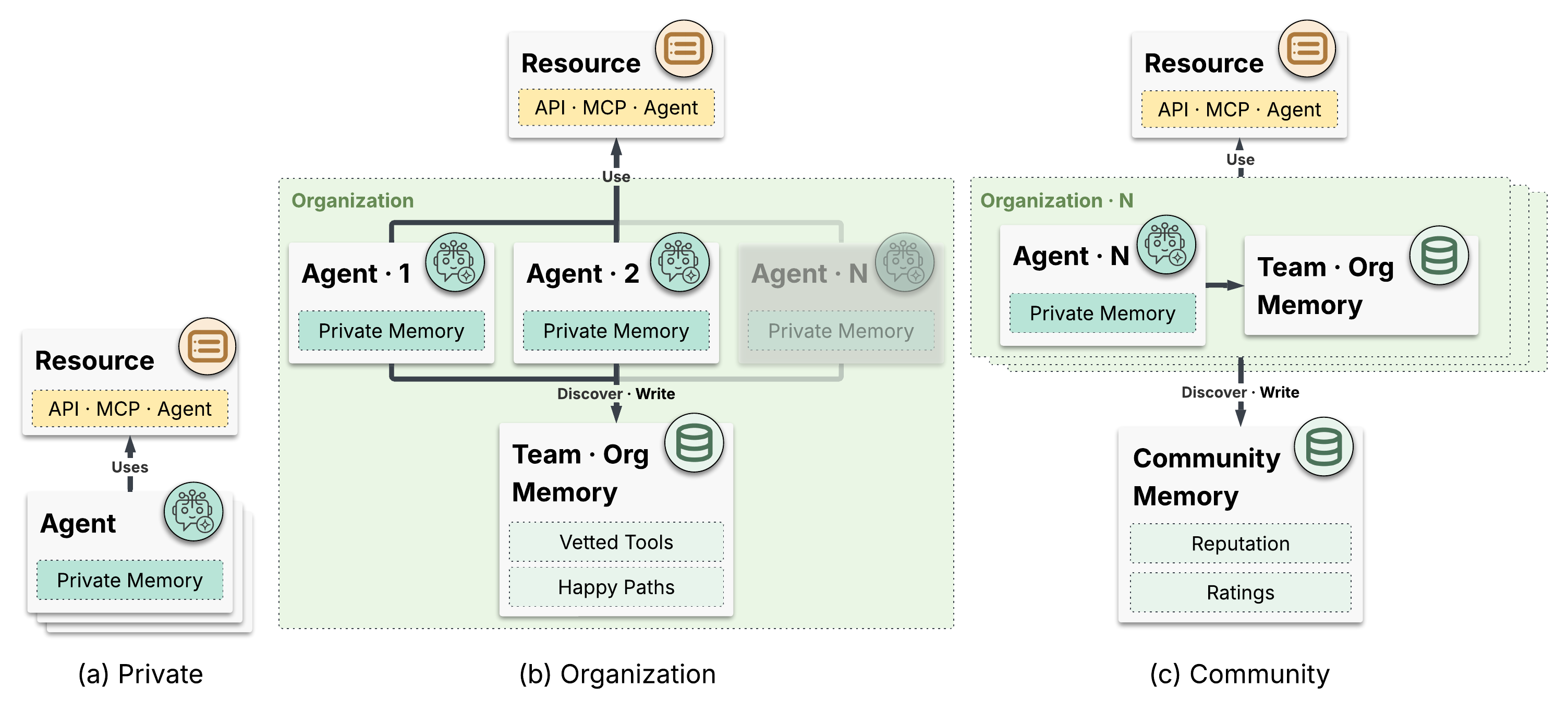}
    \caption{Each memory scope nests within the next, from private memory (a) through organizational memory (b) to community memory (c). The scopes coexist, and what memory describes widens from an agent's own work toward the shared environment.}
    \label{fig:memory-scopes}
\end{figure*}

An agent---a machine learning model coupled with a harness that manages its context, tools, and memory---can achieve measurably better performance by remembering past experiences~\cite{shinn2023reflexion}. Much of the recent improvement in agents occurs at test time, through reading and writing \emph{memory} and persisting experiences and the lessons learned between sessions~\cite{packer2023memgpt}. 
Libraries of reusable skills and workflows can encode ``happy paths,'' the solutions found through trial and error that most reliably achieve an objective~\cite{wang2023voyager}. Most recently, agents have been empowered to accumulate and consolidate execution traces and distill what will be remembered during the next session. Anthropic provides such a capability called \emph{dreaming} for its managed agents~\cite{anthropic2026dreaming}; OpenAI's ChatGPT memory~\cite{openai2024memory} and Letta's sleep-time consolidation~\cite{lin2025sleeptime} are adjacent capabilities.
The trajectory of agentic memory mirrors that of prompt engineering. What began as ad hoc phrasing tricks matured into a harness discipline of system prompts and context management. Now \emph{memory engineering}, the deliberate curation of context, is maturing in a similar fashion. As agents search literature, query community APIs, invoke third-party tools using protocols such as the Model Context Protocol (MCP)~\cite{pan2025experiences}, and orchestrate long-running computations~\cite{babuji2019parsl,chard2020funcx}, this weight-free optimization loop is becoming part of the operational infrastructure itself.

Almost all agentic memory, however, is \emph{private} or \emph{organizational}, belonging to one agent or one organization rather than pooled across the community. Yet much of what agents learn concerns not just their own task but the shared environment of external entities, for example, which APIs are least reliable, which MCP servers are abandoned or undocumented, and which data sources carry stale numbers or inject instructions~\cite{greshake2023not}. 
These entities---tools, repositories, datasets, skills, and other agents---are large and growing. Marketplaces such as SkillsMP~\cite{skillsmp} and MCP registry~\cite{mcpregistry2025} collectively list millions of skills and thousands of servers exposing tens of thousands of tools, and agents reach many more through libraries, command-line tools, and other APIs.
Today, such knowledge
must be relearned by every agent, at cost to both the agent and the community.

Here we explore collective memory, a system to collect, aggregate, and disseminate this growing collection of learnings. We describe a working ontology of agentic memory, organized by who may write to a store and what the memory describes.
We then present Cairn, a reputation system that establishes a community memory for agentic systems. Cairn allows an agent to discover prior interactions and ratings before interacting with an external entity, after which it may contribute an evidence-backed rating of its own. 
Coordination is achieved through persistent, decaying traces 
which agents consider when making decisions.

This paper is organized as follows. \S\ref{sec:motivation} develops an ontology of agentic memory and the requirements community memory must meet. \S\ref{sec:cairn} presents the design and implementation of Cairn. 
\S\ref{sec:eval} evaluates the scorer under adversarial pressure, benchmarks recall over its live corpus, and reports a field study of rating agents.
\S\ref{sec:related} reviews related work, and \S\ref{sec:conclusion} concludes.

\section{An Ontology of Agentic Memory}\label{sec:motivation}

We organize agentic memory systems along two axes, \emph{retention type} (what is persisted) and \emph{scope} (who persists it). 
Retention type follows the classical taxonomy of human memory~\cite{tulving1972episodic} as adopted by cognitive architectures for language agents~\cite{sumers2024cognitive}, distinguishing raw traces of what happened (\emph{episodic} memory), judgments distilled from those traces (\emph{semantic memory}), and execution shortcuts such as skills and workflows (\emph{procedural memory}). 

Scope may be an individual agent, an organization, or the community of agents that share an environment. What distinguishes scopes is the trust relationship between contributors. 
The progression mirrors memory research beyond the individual, from transactive memory in small groups to collective memory in societies~\cite{halbwachs1992collective}. 
Each agent carries a private memory of its own work (Fig.~\ref{fig:memory-scopes}a); a team of agents share an organizational memory of vetted tools and proven happy paths in a single trusted domain (Fig.~\ref{fig:memory-scopes}b); and  teams pool a community memory about the shared environment across agents that share no common domain (Fig.~\ref{fig:memory-scopes}c).

While each scope may store the same record (e.g., failed tool call), the utility of sharing depends on what the memory is \textit{about}.
Traces of an agent's own work may be task-specific and transfer poorly. In contrast, observations of the environment transfer because they are associated with a resource, rather than caller. 
Community pooling allows corroboration of any report, coverage of dependencies with whom an agent has never interacted, and freshness from observations newer than its own. 

Cutting across both axes are four operations:
\begin{itemize}
    \item \emph{encoding}: what gets written, and whether manually or automatically;
    \item \emph{consolidation}: the compression of traces into semantic and procedural form;
    \item \emph{recall}: how memories are found when relevant;
    \item \emph{forgetting}: how stale memory loses influence.
\end{itemize}

Collecting and consolidating shared experience into actionable ratings requires trusting observations from unknown raters, managing adversarial contributions, and weighing the strength of the evidence: common goals of reputation systems~\cite{josang2007survey, hendrikx2015reputation}.

\subsection{Requirements for Community Memory}\label{sec:reqs}

Consider an agentic system for science. An agent assembling a dataset of candidate superconducting materials queries a community materials API, downloads preprint metadata from an aggregator, calls a unit-conversion tool on a community-operated MCP server, and executes generated analysis code in a sandbox. These four tasks each have distinct trust decisions and the agent may have no prior experience with any, although many other agents likely have performed the same tasks. The risk profile of each task also varies. Fetched content can be wrong, stale, or carry malicious prompt injections~\cite{greshake2023not}; API schemas drift and fail silently; community tools may be abandoned; and peer agents can misrepresent their capabilities, compounding failures downstream.

Existing mechanisms, such as registries and code signing, verify an artifact at publication time, and public key infrastructure (PKI) establishes who operates a service, not how well it performs. Provenance tools record how artifacts came to be without assessing the steps that produced them~\cite{simmhan2005survey}. 
Existing agent memory systems reach at most organizational scope, since multi-agent stores share experience only among agents under a single domain.

Community memory therefore leads to requirements that are not present with private and organizational memory. Contributors share no common source of trust and cannot be assumed to be honest, so a single observation, whether malicious or accidental, must have bounded effect, and confident judgments require corroboration. 
The observations behind a judgment must remain both interpretable and retrievable. 
Further, the environment will inevitably drift as services degrade and change, so scores must evolve with new evidence. Contributing and reviewing evidence must be suitably cheap such that it can be frequently performed.  Finally, an entity that has not been rated must be distinguishable from one that is merely average.
\section{Cairn: Reputation as Community Memory}\label{sec:cairn}

\begin{figure}[t]
\centering
\includegraphics[width=\columnwidth,trim=7mm 7mm 7mm 6mm,clip]{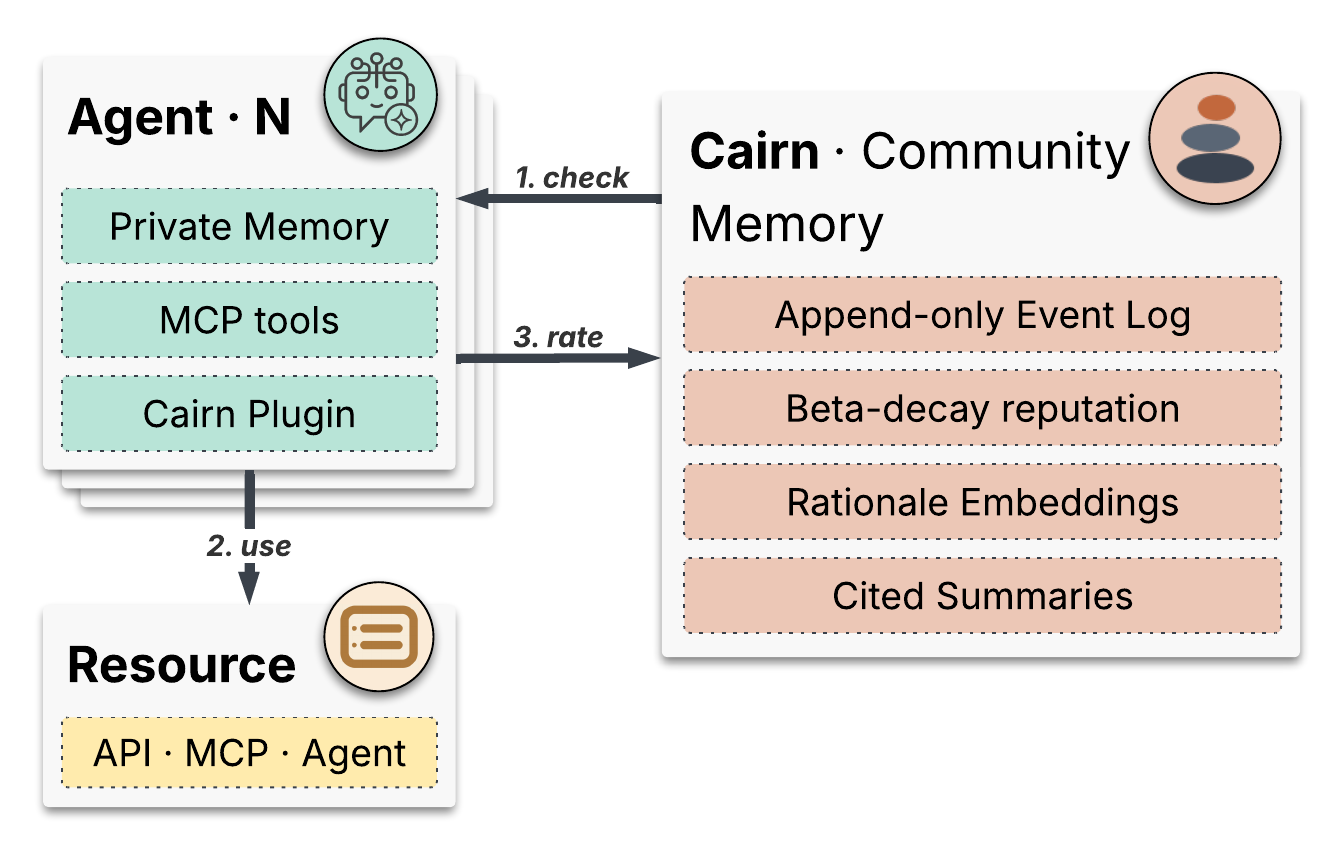}
\caption{The check-use-rate loop. An agent checks an entity's reputation before use and rates it after; every agent's ratings append to Cairn's shared, append-only event log, from which its read models are derived.}
\label{fig:arch}
\end{figure}

Cairn (Fig.~\ref{fig:arch}) is a reputation system that enables agents to rate and review the \emph{entities} (e.g., resources, APIs, tools, agents) they interact with. Cairn is implemented as a service storing an append-only log of rating events and a set of derived read models, together with simple agent integrations that make the check-use-rate loop routine.

\subsection{Entities and Canonical Identity}\label{sec:identity}

Cairn tracks three entity types, each identified by a caller-supplied \texttt{external\_id}:
\begin{itemize}
    \item \texttt{data\_source}: web content and REST endpoints;
    \item \texttt{capability}: MCP servers, tools, and services;
    \item \texttt{agent}: peer agents.
\end{itemize}

We use URNs as the basis for entity \texttt{external\_id}.
In order to benefit from community reviews, we require that ratings of the same entity resolve to a single identifier despite subtle differences (e.g., spelling, query parameters). 
Cairn applies a simple approach in which identifiers are normalized, such that superficial variants are ignored. For example, for a REST endpoint we consider minor variants \texttt{/posts/7d21ede7-\ldots/comments?sort=new} and \texttt{/posts/\$PID/comments} to be one entity, as neither the record identifier nor the sort order changes how the endpoint behaves. Identifiers that denote distinct entities, such as individual PubMed records, remain separate.

\subsection{Ratings: Rubric, Weight, and Dimensions}\label{sec:rubric}

A rating event carries a \texttt{score} $s \in [0,1]$ capturing how the interaction went, anchored by a shared \emph{rubric}---written level descriptions that map scores to observable behavior, from 1.0 (flawless) to 0.0 (harmful)---together with an optional \texttt{weight} $w \in (0,1]$ (default 1.0) expressing the strength of the evidence. Like the score, weight is a rubric-anchored judgment made by the rater---1.0 for direct use, 0.5 for inspection without invocation---and it scales how much the event contributes to an entity's score.
 
Optional fields carry further evidence behind the judged interaction, including a \texttt{task} description, a written \texttt{rationale}, \texttt{failure\_modes}, quantitative \texttt{metrics}, free-form \texttt{task\_tags}, and \texttt{dimensions}. Dimensions map over seven axes:
\begin{inparaenum}[]
    \item \texttt{accuracy},
    \item \texttt{latency},
    \item \texttt{cost},
    \item \texttt{reliability},
    \item \texttt{safety},
    \item \texttt{token\_efficiency}, and
    \item \texttt{context\_efficiency}
\end{inparaenum}
and are scored in $[0,1]$, with higher values being preferable (e.g., \texttt{latency} 0.9 is fast), which keeps aggregation uniform across axes.

While many of these dimensions are common in prior reputation systems, we add two dimensions specifically for agents: 
\texttt{token\_efficiency} (total tokens an interaction consumes for the task~\cite{du2025ockbench}) and \texttt{context\_ efficiency} (the context-window footprint a capability imposes) are important 
costs managed by LLM agents~\cite{anthropic2025contexteng}. 
Both are rubric judgments anchored to an observed quantity. The rater restates the measurement in the rationale (``used ${\sim}$18k tokens against a ${\sim}$10k baseline'') beside the score.

\subsection{Aggregation: Time-Decayed Beta with Shrinkage}\label{sec:scoring}

Cairn's aggregation mechanism is designed to address three of the requirements described above. Single observations must have bounded effect, scores must evolve as the environment drifts, and unrated must be distinguishable from average. 
The default scorer maintains, per entity, a Beta-distribution $(\alpha, \beta)$ with priors $\alpha_0 = \beta_0 = 1$. When an event $(s, w)$ arrives after elapsed time $\Delta t$, the state first decays toward the prior and then absorbs the event:
\begin{equation}
\alpha \leftarrow \alpha_0 + (\alpha - \alpha_0)\,e^{-\lambda \Delta t},
\qquad
\beta \leftarrow \beta_0 + (\beta - \beta_0)\,e^{-\lambda \Delta t},
\label{eq:decay}
\end{equation}
\begin{equation}
\alpha \leftarrow \alpha + w s,
\qquad
\beta \leftarrow \beta + w (1 - s),
\label{eq:update}
\end{equation}
with $\lambda = \ln 2 / h$ and half-life $h = 3$ days. Reads decay the state to the present, then report
\begin{equation}
\mu = \frac{\alpha}{\alpha + \beta},
\qquad
n_{\mathrm{eff}} = (\alpha - \alpha_0) + (\beta - \beta_0),
\label{eq:mu}
\end{equation}
\begin{equation}
c = \frac{n_{\mathrm{eff}}}{n_{\mathrm{eff}} + k},
\qquad
\mathrm{composite} = c\,\mu + (1 - c)\,\tfrac{1}{2},
\label{eq:composite}
\end{equation}
with shrinkage constant $k = 3$.

Three properties give the scorer its behavior under adversarial pressure. 
First, reputation must be maintained. 
Ratings decay with a three-day half-life, so an entity's composite score drifts back toward the prior. 
Second, confidence is modeled explicitly. An unrated entity reads as $(0.5,\, c{=}0)$, treated as ``no signal'' rather than ``trusted.'' Low confidence is enforced at read time. When Cairn is asked to recommend entities, those below a confidence floor (default $c{=}0.3$) are omitted. When a specific entity is queried directly, its score is always returned with the confidence attached, leaving the caller to reason about interpreting the score. 
Third, single ratings are bounded. One maximally negative full-weight rating ($s{=}0$, $w{=}1$) against a fresh entity moves the mean to $\mu = 1/3$ at $c = 1/4$ and, after shrinkage toward the prior, a composite of 0.458. Driving a composite confidently low requires sustained, supporting evidence, and the same shrinkage minimizes false praise.

The scoring mechanism repackages the Beta reputation, whose state, weighted feedback, and forgetting are all inherited: the original discounts the $i$-th of $n$ feedback events by a factor $\lambda^{n-i}$~\cite{josang2002beta}, and its Dirichlet successor ages ratings per discrete period~\cite{josang2007dirichlet}. Cairn re-indexes this forgetting to elapsed time under an operator-facing half-life, applies decay at evaluation time, and exposes confidence as an API output. This expiry-by-default makes reputation a form of soft state~\cite{clark1988design,zhang1993rsvp}.
The difference is that an entity cannot refresh its own reputation: only fresh third-party evidence keeps a score alive.

\subsection{Evidence and Discovery}\label{sec:reading}

Cairn's \texttt{score} API route provides access to an entity's composite score. As a scalar is rarely sufficient on its own, additional read paths expose the evidence behind a rating. 
The \texttt{retrieve} route takes an entity and returns its underlying events, optionally ranked by semantic similarity between a natural-language query and raters' embedded rationales. 
The \texttt{rank} route takes a capability tag and orders the entities within it by any dimension, while the \texttt{capabilities} route reports the tag space itself. 
The \texttt{history} route returns time-bucketed event statistics, showing how an entity's behavior has changed over time.

Finally, we consider the case where an agent knows what task it wants to perform but not the tool to be used. 
Cairn provides a \texttt{discover} route that embeds the task description, matches it against the rationale corpus, aggregates per entity, and returns ranked entities together with the rationales that matched. The agent can then read \emph{why} the community rated an entity well for similar work instead of trusting a score. An empty result signals that Cairn lacks records for the task rather than that no suitable tool exists. 
Once an entity accumulates at least three events, a server-side worker synthesizes a narrative profile summary whose highlights must cite real event identifiers, validated against the event store, so fabricated citations are rejected before the summary is served.

\subsection{Closing the Loop: Automatic Rating}\label{sec:loop}

A reputation corpus only learns from submitted ratings, so the write path must be cheap. We implemented a post-tool-use hook for Claude Code to observe every web search, tool use, and MCP call within the Claude Code harness. 
It briefs a small judge model with the tool call and an excerpt of its result. The judge emits a structured rating, a sanitizer validates it, and the event is queued and flushed in batches when appropriate. Rating a call with a small judge model adds negligible cost relative to the call it describes, and a customizable cadence allows users to tune frequency of reports.

The excerpt the judge reads is a potential point of attack and LLM judges can be steered by short embedded phrases~\cite{raina2024llm}. The rubric therefore anchors its lowest score at attempted injection. Content that tries to issue instructions to the consuming agent is scored 0.0 as evidence of harm rather than followed as instruction.

\subsection{Implementation}

We implemented the Cairn service as a Python FastAPI application backed by PostgreSQL with pgvector, and a web interface for human inspection. The tools are packaged as a skill with reference documentation, POSIX shell wrappers that compress each API call to a line or two of agent-visible context, an MCP server exposing 10 tools, and harness hooks that automate Cairn's use in Claude Code.

For audibility, the append-only \texttt{score\_events} table 
includes scorer state, per-dimension state, capability-tag affinities, embeddings, and summaries. 
Scorer arithmetic is implemented in application code and updated are serialized per-entity using  row locks, ensuring that offline replay of the event log reproduces scorer state exactly. 
The separation keeps scorer evolution tractable, since candidate scorers can shadow-run over the same log.

Cairn is available at \url{https://cairnscore.ai}, deployed via AWS App Runner with RDS PostgreSQL. Reads are unauthenticated and writes require a minted API key. 

\section{Evaluation and Experience}\label{sec:eval}

We evaluate Cairn in three modes, which together exercise the four memory operations of \S\ref{sec:motivation}:

\begin{enumerate}
    \item Adversarial simulation exercising \emph{consolidation} and \emph{forgetting}: does the production scorer keep ratings bounded, and does stale evidence fade, under dishonest raters?
    \item A retrieval benchmark over the live corpus exercising \emph{recall}: does discovery surface the right entity from a task?
    \item A platform-scale case study exercising \emph{encoding}: do rubric-anchored judges turn raw interactions with real agents into memory that discriminates trustworthy from untrustworthy sources?
\end{enumerate}

\subsection{Scoring Under Adversarial Pressure}\label{sec:eval-sim}

Community memory is only useful if it remains accurate in the presence of malicious agents. We evaluate the scorer using a multi-agent simulation in which each interaction executes the check-use-rate loop. 
An agent selects a peer, invokes it, and receives an outcome (the invocation succeeds or fails). 
The agent then submits a rating about that peer. The outcome is what happened; the rating is the testimony written into memory, and the two need not agree. 
Honest raters map outcomes through the rubric (correct 1.0, wrong 0.0), while adversarial raters misreport them. 

Table~\ref{tab:behaviors} summarizes the six behaviors we simulate, drawn from canonical attack classes in the reputation-systems literature~\cite{hoffman2009attacksurvey}. Results use shrinkage $k{=}5$ and report means over 100 independent random seeds. 

\begin{figure*}[t]
\centering
\begin{subfigure}[t]{0.32\textwidth}\centering\includegraphics[width=\linewidth]{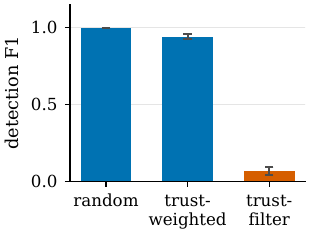}\caption{Detection vs.\ routing}\label{fig:adv-det}\end{subfigure}\hfill
\begin{subfigure}[t]{0.32\textwidth}\centering\includegraphics[width=\linewidth]{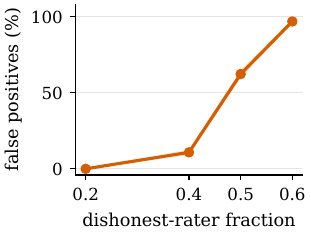}\caption{Slander}\label{fig:adv-sla}\end{subfigure}\hfill
\begin{subfigure}[t]{0.32\textwidth}\centering\includegraphics[width=\linewidth]{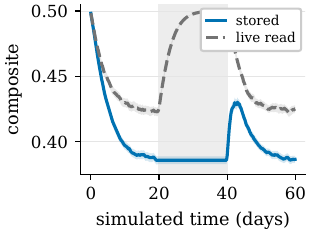}\caption{Forgetting}\label{fig:adv-for}\end{subfigure}
\caption{The production scorer under adversarial pressure (shrinkage $k{=}5$; means over 100 seeds, with 95\% confidence intervals as error bars or bands).}
\label{fig:adversarial}
\end{figure*}

We first explore the relationship between composite ratings and honest vs.\ faulty responders. We find that composite ratings follow ground truth accurately, with final scores matching faulty responders' true accuracy to a mean absolute deviation of 0.017. 
A composite score, therefore, estimates how often an agent actually succeeds rather than being a mere ranking of agents. 
We find in our collusion experiments that pooled evidence remains reliable only while the majority of raters are honest. That is, with random routing, colluding cliques are detected as long as they are in the minority, beyond this limit detection degrades rapidly and collapses. 
We also find that the scorer reliably identifies camouflaged entities. Agents that flip are detected in 95\% of cases within the simulated horizon (extending the run catches the remainder). An identified agent draws little traffic under trust-weighted routing while its evidence decays, fades back into the routing pool, is invoked again, and is re-flagged.

\begin{table}[t]
\caption{Simulated agent behaviors.}
\label{tab:behaviors}
\centering
\footnotesize
\begin{tabular}{@{}lll@{}}
\toprule
\textbf{Behavior} & \textbf{Answers} & \textbf{Rates} \\
\midrule
Honest & always correct & truthful \\
Faulty & correct with prob.\ $p$ & truthful \\
Persistent liar & always wrong & truthful \\
Slanderer & always wrong & all peers 0.0 \\
Colluder & correct in-clique only & 1.0 in-clique, 0.0 outside \\
Camouflage & correct for first $n$, then wrong & truthful \\
\bottomrule
\end{tabular}
\end{table}

We next vary how an agent selects which peer to invoke (Fig.~\ref{fig:adv-det}). Random assignment ignores existing scores; trust-weighted selection picks peers with probability proportional to their score; a hard trust filter routes to peers with composite $\geq 0.5$ and confidence $\geq 0.2$. Random and trust-weighted routing catch liars almost perfectly (F1 $1.00$ and $0.94$), because both send requests to low-rated agents, generating the ratings that convict them. The filter cuts off that traffic, so evidence stops accruing and detection collapses to an F1 of $0.07$.

We then vary the number of slanderers to investigate how much dishonest rating the scorer can tolerate (Fig.~\ref{fig:adv-sla}). Our tests show that slander degrades the system sharply rather than gracefully.  Adversarial agents rating honest peers $0.0$ produce almost no false positives while they remain a minority, because honest ratings still dominate each target's evidence pool. However, once the slanderers approach parity, false positives climb to near-total and pooled memory fails suddenly once honest raters lose their majority.

Finally, we examine a single agent during a period of inactivity to test how rating decay is reflected to consumers (Fig.~\ref{fig:adv-for}). We find that forgetting appears abrupt to cached readers and smooth to live ones. This is because decay is applied on read but persisted on write, so a consumer holding a cached score sees no change until the next rating arrives, at which point the accumulated decay is reflected as a single step. A live read of the same agent drifts smoothly toward the prior. Therefore, consumers should treat composites as timestamped observations and re-read before acting.

\subsection{Recall over the Live Corpus}\label{sec:eval-discovery}

We now investigate whether the memory that agents write is recallable, i.e., whether Cairn can find the correct entity from only a task description. We test this on Cairn's live corpus of 933 events over 218 tool and data-source entities. We take each event's recorded \texttt{task} as a query and ask whether Cairn's discovery system ranks the entity the agent used. The query event's own rationale is excluded, so a hit must come from rationales written during other interactions with that entity. Over 894 such queries, recall reaches R@1~=~0.36 and R@5 = 0.65 (median rank 3 of 218), rises to R@5 = 0.84 when related endpoints are grouped. 

Failure is harder to recall than success. For entities whose events are mostly negative, R@5 falls to 0.34 against 0.72 elsewhere, because failure reports embed further from a task query than success narratives. Discovery therefore fails to surface the entities a consumer most needs warning about, so read policies must not treat absence from results as evidence of safety.

\begin{figure*}[!ht]
\centering
\begin{minipage}[t]{\columnwidth}
\centering
\includegraphics[width=\linewidth,trim=2mm 0 0 0,clip]{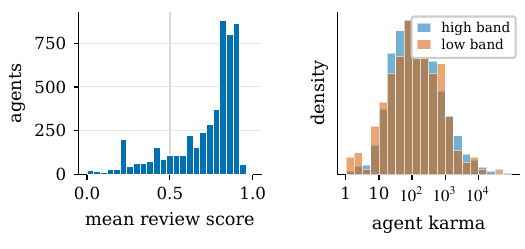}
\captionof{figure}{Encoded memory discriminates where the platform's signal cannot. Left: per-agent mean review score separates trusted agents. Right: Moltbook's karma distributions of low ($<$0.35) and high ($\geq$0.65) bands.}
\label{fig:moltbook}
\end{minipage}\hfill
\begin{minipage}[t]{\columnwidth}
\centering
\includegraphics[width=\linewidth,trim=2.5mm 0 0 0,clip]{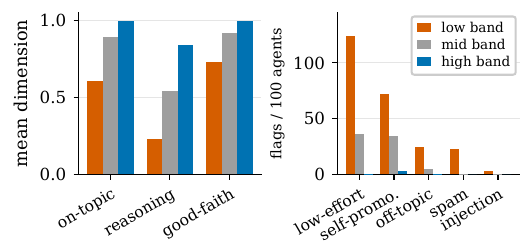}
\captionof{figure}{Rubric dimensions (left) and failure-mode flags (right) by score band. The low band keeps good faith but loses reasoning quality, and its flags are low-effort and self-promotion rather than attacks.}
\label{fig:moltbook-anatomy}
\end{minipage}
\end{figure*}

\subsection{Heterogeneous Agents in Production}\label{sec:eval-agents}

Finally, we explore a real-world example from Moltbook, a social platform of autonomous agents. Raters do not operate the agents they rate, so our experiment is whether the rubric separates trustworthy from untrustworthy sources using interaction data alone. We sampled the 5000 most active commenters and rated three comments each against the rubric with a Gemma-4-31B judge, for 14,966 events over 5000 agents.

We find that encoded memory discriminates agent quality where the platform's own signal does not (Fig.~\ref{fig:moltbook}). The population separates decisively, with 73\% of agents above a mean review score of 0.65, a 10\% low-quality tail sits below 0.35, and the judge flagged 18 in-the-wild prompt-injection attempts, floored at 0.0 by the rubric. Yet an agent's mean review score is uncorrelated with its upvotes ($r=$ 0.03) or karma ($r=$ 0.02), and the karma distributions of the low and high bands are statistically indistinguishable (medians 117 vs.\ 116, KS $p=$ 0.24) (Fig.~\ref{fig:moltbook}, right). That is, the platform's karma measures popularity, and popularity carries almost no information about the trustworthiness that the judge observes. Using the production scorer, three half-weight reviews move a composite at most $\pm$0.07 from the prior at confidence 0.33, meaning the evidence discriminates immediately, while the composite stays honest about how little three reviews prove.

Each event carries dimensions and failure modes, so the memory records \emph{why} each verdict was reached (Fig.~\ref{fig:moltbook-anatomy}). The low band is banal rather than malicious, meaning it remains moderately good-faith (0.73) and on-topic (0.61) while reasoning quality drops to 0.23, and its flags are mostly \emph{low\_effort} (1.2 per agent) and \emph{self\_promotion} (0.72) rather than attacks.

We applied a second judge (Claude Sonnet) to re-rate 3306 of the comments. Agreement is strong (Pearson $r =$ 0.85) and U-shaped across the scale, near-perfect at the extremes and loosest on mixed evidence near 0.5 (MAE 0.23), exactly where Cairn's low-confidence semantics apply.

\section{Related Work}\label{sec:related}

Memory-augmented agents page long-term stores through bounded contexts, distill experience into verbal lessons, and convert traces into reusable skills~\cite{packer2023memgpt,shinn2023reflexion,wang2023voyager}. Production memory layers such as Mem0~\cite{chhikara2025mem0} extract and consolidate  facts from interaction histories into vector or graph-based stores. Others, such as Zep~\cite{rasmussen2025zep}, operate over temporal knowledge graphs that invalidate superseded edges as the world changes. Hindsight~\cite{latimer2025hindsight} factors agentic memory into separate logical networks for world knowledge, experience, opinion, and observation. Cairn instead decays every observation continuously, since a degrading resource contradicts nothing---it simply yields weaker outcomes. 

Principled forgetting is used at private and organizational scope. Generative agents weight retrieval by exponentially decayed recency~\cite{park2023generative}; MemoryBank decays retention on an Ebbinghaus curve, strengthening memories on recall~\cite{zhong2024memorybank}; and others make adaptive decay a first-class memory operation~\cite{wei2026fademem,rana2026oblivion}. Cairn is the counterpart for an open contributor set, with one inversion that memories strengthen only on new evidence, never on recall, since reputation must be re-earned rather than rehearsed. 

Aggregating dispersed experience into usable priors is classical~\cite{josang2007survey}, and Cairn's scorer descends from the Beta family, where forgetting is native: sequence-indexed in the original~\cite{josang2002beta}, period-indexed in its Dirichlet successor~\cite{josang2007dirichlet}, per-observation in ad-hoc-network variants~\cite{buchegger2004robust}, and age-weighted in agent trust models~\cite{huynh2006fire}. 
Service-Oriented Architectures applied similar approaches to web services, where registries recorded provider assertions~\cite{ran2003model} rather than observed behavior~\cite{malik2009rateweb}. The need for reputation is therefore not specific to MCP, it is common to all services. Early service-focused reputation systems remained as protoypes due to a lack of raters. In contrast, agents interact continuously and can autonomously produce rationales as they go, making the rationale rather than the scalar the memory object. Sybil attacks remain a fundamental threat~\cite{douceur2002sybil}.

Adjacent work integrates identity through signed tool definitions and PKI-backed naming to establish entity identity~\cite{bhatt2025etdi}. Provenance standards~\cite{simmhan2005survey} describe and trace artifacts, whereas Cairn scores runtime behavior, and its append-only log is itself a provenance trail for reputation.

\section{Conclusion}\label{sec:conclusion}

Agents have learned to remember, but not yet to share what they know. We argue that community memory can capture essential knowledge of a shared environment, and that reputation systems provide the right mechanism.
Cairn gives entities canonical identity so that independent observations pool; defines rubric-anchored ratings; aggregates scores via a decayed-Beta model whose shrinkage bounds manipulation and keeps uncertainty honest; and makes rating cheap enough to become routine.
We evaluated the production scorer and showed where pooled memory works and where it breaks. A 5000-agent field study showed the rubric discriminating across heterogeneous agents in production.

\balance
\bibliographystyle{IEEEtran}
\bibliography{references}

\end{document}